\documentclass[aps, preprintnumbers, nofootinbib, floatfix, amsmath, amssymb, runinaddress, tightenlines, 11pt]{revtex4}
\pdfoutput=1

\usepackage{amssymb,amsmath,amsfonts,amsbsy,epsfig,wrapfig,caption,float,slashed,graphicx, pstricks}

\newcommand{\eq}[2]{\begin{equation}\label{#1}#2 \end{equation}}

\usepackage{xcolor}

\newcommand{\p}{\varphi}
\newcommand{\mpl}{M_{\rm pl}}

\newcommand{\calL}{{\cal L}}
\newcommand{\calP}{{\cal P}}
\newcommand{\calR}{{\cal R}}

\newcommand{\calA}{{\cal A}}

\newcommand{\w}{\widetilde}

\def\bea{\begin{eqnarray}}
\def\eea{\end{eqnarray}}
\def\be{\begin{equation}}
\def\ee{\end{equation}}
\def\ba{\begin{array}}
\def\ea{\end{array}}

\def\nn{\nonumber}

\begin{document}
\preprint{CERN-PH-TH-2014-205}
\author{Ignatios Antoniadis$^{1,2}$}
\email{Ignatios.Antoniadis@cpht.polytechnique.fr}
\author{Subodh P. Patil$^{3}$}
\email{subodh.patil@cern.ch}
\affiliation{1) Albert Einstein Center for Fundamental Physics, Institute for Theoretical Physics,
Bern University, Sidlerstrasse 5 CH-3012 Bern, Switzerland\\} 
\affiliation{2) Ecole Polytechnique, 91128 Palaiseau, France\\}
\affiliation{\\3) Theory Division, PH-TH Case C01600, CERN, CH-1211 Geneva, Switzerland\\}
\date{\today}

\title{The Effective Planck Mass and the Scale of Inflation}

\begin{abstract}
Observable quantities in cosmology are dimensionless, and therefore independent of the units in which they are measured. This is true of all physical quantities associated with the primordial perturbations that source cosmic microwave background anisotropies such as their amplitude and spectral properties. However, if one were to try and \textit{infer} an absolute energy scale for inflation-- a priori, one of the more immediate corollaries of detecting primordial tensor modes-- one necessarily makes reference to a particular choice of units, the natural choice for which is Planck units. 
In this note, we discuss various aspects of how inferring the energy scale of inflation is complicated by the fact that the effective strength of gravity as seen by inflationary quanta necessarily differs from that seen by gravitational experiments at presently accessible scales. The uncertainty in the former relative to the latter has to do with the unknown spectrum of universally coupled particles between laboratory scales and the putative scale of inflation. These intermediate particles could be in hidden as well as visible sectors or could also be associated with Kaluza-Klein resonances associated with a compactification scale \textit{below} the scale of inflation. We discuss various implications for cosmological observables. 
\end{abstract}

\maketitle

\section{Preliminaries}
The strength of the gravitational force depends on the scale at which it is measured\footnote{e.g. via Cavendish type experiments where we have precise knowledge of two masses (one of which could be a test mass), or equivalently in principle through gravitational scattering experiments.}. At laboratory scales, the strength of gravity is characterized by the reduced Planck mass $\mpl = 2.435\times 10^{18}$ GeV which determines Newton's constant $G_N = \mpl^{-2}$. However, like all other interactions, quantum corrections effect the effective strength of gravity depending on the characteristic energy of the process probing it\footnote{See \cite{Donoghue, Burgess} for reviews of treating gravity as an effective theory. In the following discussion, we steer clear of potentially problematic aspects of the notion of running gravitational couplings \cite{running} by focussing only on physically observable quantities such as amplitudes and cross sections.}. 

Massive particles are particularly interesting for the threshold effects they impart once we start to probe energies above their mass $M$, i.e. at distances below $M^{-1}$. This can be understood via a simple thought experiment \cite{Dvali1}: consider scattering a test particle off a very heavy point mass. The inverse Fourier transform of the scattering amplitude yields the gravitational potential generated by the source. Once the inter-particle separation approaches $\Delta x \sim M^{-1}$, $M$ being the mass of some heavy particle, virtual pairs of these particles are created, the positive/ negative energy virtual quanta of which are attracted/ repelled by the source, creating a gravitational dipole distribution that effectively anti-screens the source, strengthening its gravitational field. Therefore, the strength of gravity is increased by this effective `vacuum polarization' far enough away from the threshold induced by a particle of mass $M_j$ that couples to gravity\footnote{This is true regardless of whether these massive particles couple directly to the sector that contains the probe particle (e.g. the Standard Model) or not.}, i.e. as we probe increasingly shorter distances $\Delta x \ll M^{-1}_j$.

One can quantitatively understand this effective strengthening through the computation of the graviton propagator with loops of the massive fields contributing to the graviton self-energy insertions. We trace through the details of this computation in appendix \ref{appendix} following the treatment of \cite{capper}, however a quick understanding of this can be arrived at through the argument presented in \cite{Dvali2}. Consider the correction to the graviton propagator induced by loops of various particles-- suppressing all index structure, we find that the leading correction will have the form
\eq{propcorr}{\frac{1}{\mpl^4}\frac{1}{p^2}\langle T(-p) T(p) \rangle \frac{1}{p^2},}
where $T(p)$ schematically represents the \textit{total} energy momentum tensor of the theory. We further consider the limit where the external momentum satisfies $p^2 \gg M^2$ where $M^2$ is the mass of the heaviest particle that can run through the loops. In this limit, the theory becomes conformal, which fixes the finite part of the loop integral to be
\eq{climit}{\langle T(-p) T(p) \rangle \sim \frac{c}{16\pi^2} p^4\,{\rm log}\frac{p^2}{\mu^2}} 
where $\mu$ is some arbitrary renormalization scale, and where $c$ is the central charge of the theory that effectively counts the number of degrees of freedom running through the loop, i.e. $c \approx N$ \cite{Dvali2}. As an illustrative example, in four dimensional Minkowski space, the central charge of a non-interacting theory containing $N_\phi$ scalars, $N_{\psi}$ Dirac fermions and $N_V$ vector bosons is given by \cite{Duff}\cite{Osborn}\cite{ratt}: 
\eq{ccdef}{c := \w N = \frac{4}{3}N_\phi + 8N_\psi + 16 N_V.}
Comparison with the free propagator $1/(p^2\mpl^2)$ implies that the perturbative expansion fails at the scale $p = M_{**}$, where 
\eq{pexp}{M_{**} \sim \frac{\mpl}{\sqrt{\w N}},}
which is when gravity becomes strongly coupled. \textit{That is, $M_{**}$ is the effective cut-off of gravity at short distances}. However one must take care to distinguish between the scale of strong gravity $M_{**}$ from the strength of gravity at a particular energy scale, which we denote $M_*$. Whereas the former sets the scale at which unitarity starts to break down in the effective theory\footnote{And is thus unitarized by the appearance of new degrees of freedom at $M_{**}$ from some UV completion, such as string theory.}, the latter determines the strength of gravitationally mediated processes at any particular scale \textit{below} $M_{**}$. As detailed in appendix \ref{note}, although every massive species contributes to lowering the scale at which strong gravity effects become important, one has to distinguish between species that universally couple directly to the matter energy-momentum tensor at tree level (such as massive Kaluza-Klein (KK) gravitons,  non-minimally coupled scalars and $U(1)$ gauge fields) from ordinary four dimensional fields that couple at one loop, in terms of their effects on the strength of gravity \textit{as one crosses the threshold set by the mass $M$ of the species,} but are still far below the scale $M_{**}$. Whereas the former immediately effect the strength of gravity, the latter do not make their effects known until very close to $M_{**}$\footnote{We are grateful to Sergey Sibiryakov for discussions concerning this point.}. Therefore, for the rest of our discussion we denote $N$ to be shorthand for the weighted index that effectively counts the number of universally coupled degrees of freedom below the energy scale of interest corresponding to the generalization of (\ref{ccdef}), such that the strength of gravity at that scale, henceforth taken to be the scale of inflation, is given by
\eq{strength}{M_* \sim \frac{\mpl}{\sqrt N}}
In what follows, we work out the consequences of this scale dependence of of the strength of gravity for inferring various quantities during inflation, which we take to be driven by a single field for economy of discussion and because the data doesn't compel us to consider otherwise \cite{planckcos, planckinf}\footnote{See also the related studies \cite{scaleinf} that explore how additional fields and non-adiabaticity further complicates inferring the scale of inflation from the detection of primordial tensors.}. As is to be expected, all dimensionless observables such as the amplitude and spectral properties of the perturbations are unaffected by the changing strength of gravity at inflationary energies. However, when one tries to \textit{infer} an absolute energy scale for inflation, one finds that it is undetermined commensurate with (\ref{strength}) up to the unknown spectrum of universally coupled species between laboratory scales and the inflationary scale, the details of which we elaborate upon in the following.

\section{The scale of Inflation}
According to the inflationary paradigm, the primordial perturbations observed in the CMB were created at horizon crossing during the quasi de Sitter (dS) phase of early accelerated expansion sourced by the inflaton field. Therefore all quantities that enter calculations of primordial correlation functions (which we subsequently relate to observables in the CMB) refer to quantities at the scale at which inflation occurred. We denote all quantities measured at the scale of inflation with a starred subscript. The dominant contribution to the temperature anisotropies comes from adiabatic perturbations \footnote{In what follows, we assume that all of the extra species have sufficiently suppressed couplings to the inflaton during inflation (e.g. either through derivative couplings or as Planck suppressed interactions) so that isocurvature perturbations are not significantly generated. This is trivially true for hidden sector fields.} 
sourced by the comoving curvature perturbation $\calR$, defined as the conformal factor of the 3-metric $h_{ij}$ in comoving gauge:\footnote{Comoving (or unitary) gauge is defined as the foliation where inflaton field fluctuations have been locally gauged away. In words, it is the time slicing where the inflaton itself is the clock.}
\eq{indef}{h_{ij}(t,x) = a^2(t)e^{2\calR(t,x)}\hat h_{ij};~~~ \hat h_{ij} := {\rm exp}[\gamma_{ij}]}  
with $\partial_i \gamma_{ij} = \gamma_{ii} = 0$ defining transverse traceless graviton perturbations. The temperature anisotropies are characterized by the dimensionless power spectrum for $\calR$, whose amplitude is given by  
\eq{prdef}{\calP_\calR := \frac{H^2_*}{8\pi^2 M_*^2 \epsilon_*} = \calA \times 10^{-10},}
where $\epsilon_* := -\dot H_*/H_*^2$, $H_*$ being the Hubble factor during inflation. Given that $\calR$ is conserved on super-horizon scales (in the absence of entropy perturbations), this immediately relates to the amplitude of the late time CMB anisotropies, which fixes $\calA \sim 22.15$ \cite{planckcos}. The tensor anisotropies are characterized by the tensor power spectrum
\eq{ptdef}{\calP_\gamma := 2\frac{H^2_*}{\pi^2 M_*^2},}
Taking the ratio of the above with (\ref{prdef}), we find the tensor to scalar ratio
\eq{str}{r_*:= \frac{\calP_\gamma}{\calP_\calR} = 16\epsilon_*.}
Therefore any determination of $r_*$, either through direct measurements of the stochastic background of primordial gravitational waves or through their secondary effects on the polarization of the CMB \cite{polnarev, Kamionkowski:1996zd, Seljak:1996gy} allows us in principle to fix the scale of inflation. Specifically, by re-expressing (\ref{prdef}) as
\eq{Hinf}{H_* = M_* \left(\frac{\pi^2\calA r_*}{2\cdot 10^{10}}\right)^{1/2},}
one can determine the value of the potential during inflation in the slow roll approximation:
\eq{Vinf}{V_*^{1/4} = M_*\left(\frac{3\pi^2\calA r_*}{2 \cdot 10^{10}}\right)^{1/4}}
We see that any measurements of $r_*$ and $\calA$ determines the scale of inflation \textit{up to our ignorance of the effective strength of gravity at the scale $H_*$}, given by
\eq{rmp}{M_* \sim \frac{\mpl}{\sqrt N}}
where $\mpl = 2.435\times 10^{18}$ GeV is the reduced Planck mass that defines the strength of gravitational interactions at laboratory scale wavelengths and longer. As noted above, $N$ is a weighted index that effectively counts the number of all universally coupled\footnote{So that (\ref{rmp}) denotes a tree level relation.} species up to the scale $H_*$-- whether they exist in the visible sector or in any hidden sector. Presuming $r_* = 0.1$, (\ref{Vinf}) implies an energy scale for inflation of $V_*^{1/4} = 7.6\times 10^{-3}M_*$. 

In order to keep track of concepts in the discussion to follow, we distinguish between what we henceforth refer to as \textit{the scale of inflation}-- defined as $H_*$ during inflation-- and the \textit{energy scale of inflation}, defined as $V_*^{1/4}$. The reason for this distinction is that $H_*$ defines (among other things) the scale above or below which massive particles respond to the background expansion irrespective of any direct couplings to the inflaton\footnote{In addition to quantum corrections to the effective action itself being set by the ratio $H^2_*/M^2_*$.} whereas $V^{1/4}_{*}$ defines the physical energy density in the inflaton field as seen by particles that couple to it, such as all decay products produced in (p)reheating. We take this distinction for granted in what follows.

In a universe where there is a true desert between laboratory scales and the onset of inflation\footnote{Note that this would require a desert not only in the sector in which the standard model resides, but in all other hidden sectors as well.}, $M_* = \mpl$. 
However, given our ignorance of particle physics between collider scales and the scale of inflation in addition to all hidden sector physics, $M_*$ is in general lower than $\mpl$ according to (\ref{rmp}), where $N$ represents a model dependent parameter that obscures our ability to infer the actual energy scale of inflation from observations of CMB temperature and polarization anisotropies. That is:
\eq{vact}{\boxed{V_*^{1/4} \sim \frac{r_*^{1/4}}{\sqrt N}\,3.28 \times 10^{16}\,\rm{GeV}.}}
Presuming a range for $r_*$ such that $0.001 \lesssim r_* \lesssim 0.1$, it is amusing to infer that in order to have an energy scale of inflation around 10 TeV, one requires $N \sim 10^{26}$ universally coupled species directly to the matter stress-tensor with masses less than that energy. Presumably any such particles in the visible sector would have started to appear in collider events accessed at the LHC. Note that as one lowers the scale of strong gravity, the maximum reheating temperature $T_i$ is necessarily lowered as well, since it cannot be higher than (\ref{vact}). Conservatively, $T_i$ cannot be too far below the TeV scale without spoiling the standard scenarios of big bang cosmology-- in particular, mechanisms for Leptogenesis and Baryogenesis which can occur no lower than the electroweak scale \cite{EW}.

We note as a consistency check on the above considerations, that although additional species increase the strength of gravity, the ratio $H_*^2/M_*^2$ is independent of $N$ and is fixed by observable quantities as
\eq{ratio}{\boxed{\frac{H_*}{M_*} = \left(\frac{\pi^2\calA r_*}{2\cdot 10^{10}}\right)^{1/2} := 
\Upsilon = 1.05\,\sqrt r_* \times 10^{-4}}}
Therefore the effects of strong gravity are evidently negligible during inflation even if $M_*$ is much smaller than the macroscopic strength of gravity $\mpl$. Hence inflationary dynamics, in particular the dynamics of adiabatic fluctuations remain weakly coupled independent of $N$ and the usual computation of adiabatic correlators can be implemented \cite{Cheung:2007st}.

\section{Extra species as Kaluza-Klein states}
It is an interesting exercise to work out the consequences of extra species associated with the Kaluza-Klein (KK) states of a particular compactification. One of two scenarios are possible-- that inflation occurs below ($H_* < \mu_c$), or above ($H_* > \mu_c$) the effective compactification scale $\mu_c$ defined as the mass scale associated with the moduli that fix the size of the extra dimensions\footnote{We presume for simplicity that there are no further hierarchies between the extra dimensions. Note that $\mu_c$ can be in general (hierarchically) different from the actual compactification scale associated to their inverse size.}. In the former case, the moduli corresponding to the extra dimensions remain fixed at their minima during inflation and we can avail of the usual relation between the fundamental gravity scale $M_{**}$ \textit{below the effective compactification scale} and the long wavelength strength of gravity (the Planck mass):
$$M_{**}^{D-2}V_n = \mpl^2\,,$$
where $V_n$ is the volume of the compactified sub-manifold \cite{aadd}. Again, the double asterisked subscript is to differentiate $M_{**}$ from $M_*$, the strength of gravity at the inflationary scale $H_*$. In $D = 4 + n$, this relation becomes $M_{**}^{2 + n}V_n = \mpl^2$. For the example of toroidal compactifications\footnote{This will remain to be true for more general compactifications (up to factors of order unity) provided again that there are no further hierarchies among the extra dimensions.}, $V_n = M^{-n}$ so that
\eq{usual}{M_{**}^{2}\frac{M_{**}^n}{M^n} := M_{**}^2V_{**} = \mpl^2\,,}
where we have defined $V_{**}$ as the volume in units of $M_{**}$. Comparison with (\ref{pexp}) implies
\eq{nvol}{\widetilde N = V_{**},}
where we again distinguish $\widetilde N$ from $N$, the former of course being the total number of species up to the effective cut-off whereas the latter is the total number of species up to the scale $H_*$. To see this another way, we note that we could also have arrived at (\ref{nvol}) through more direct reasoning. Consider first for simplicity a tower of KK states on a single flat, compact dimension of radius $R = M^{-1}$. The KK modes are characterized by their quantized momenta along the extra dimension, resulting in a tower of masses:
\eq{kkm}{m_l^2 = l^2M^2\quad l=0,\pm1,\dots.}
Clearly, the maximum number of resonances there can be before one hits the scale of strong gravity is set by the highest permissible momentum quantum number $l^2_{max}M^2 = M^2_{**}$. Hence with one extra dimension, the number of extra massive species is given by $\widetilde N = M_{**}/M$. With $n$ extra dimensions, we have $m_{\{l_i\}}^2 = \sum_i l_i^2 M^2$ with $i$ running from $1$ to $n$. The number of extra massive species is now given by (neglecting factors of order unity): 
\eq{Nvol}{\widetilde N = (M_{**}/M)^n \equiv V_{**},}
which corresponds to the number or lattice sites such that the condition $\sum_i l_i^2 M^2 \leq M_{**}^2$ is satisfied. We note that one could also have inflation happen above the effective compactification scale ($H_* > \mu_c$). In general, this would involve having to track the full dynamics of the moduli fields on their way to stabilization which does not permit any straightforward generalizations. However there are certain limits for which the moduli are effectively frozen in spite of not being fixed at the minima of their effective potentials. As discussed in appendix \ref{moduli}, this occurs in the limit where the sum of the inflationary and moduli potentials satisfy an analog of the slow roll conditions. We will presume this to be the case when $H_* > \mu_c$. Although the discussion to follow presumes $H_* < \mu_c$, the results generalize straightforwardly were we to replace $M_{**}$ with $\bar M_{**}$ defined as the effective cut-off when the compact dimensions have the (effectively frozen) volume $\bar V_{**}$ during inflation.

\subsection{Extra KK species and the scale of inflation}
During inflation, all masses much lighter than $H_*$ correct the graviton propagator and will contribute towards lowering the effective gravitational cut-off. If furthermore, these states are universally coupled (as are KK gravitons), they will also increase the effective strength of gravity now set by $M_*$.  All heavier KK states do not correct the short range interactions (i.e. they decouple) and can safely be ignored. Therefore $N$ the number of massive species that correct the strength of gravity is bounded by $n^2 M^2 = m_n^2 \ll H_*^2$ in the case of one extra dimension. Hence
\eq{Nbound}{N \lesssim \frac{H_*}{M}.}
Imagine we were to saturate this bound-- 
\eq{Nbound}{N \approx \frac{H_*}{M_*}\frac{M_*}{M} \approx \Upsilon \frac{M_*}{M} \approx 1.05\,\sqrt r_* \times 10^{-4}\frac{M_*}{M}}
where the latter follows from the observationally determined quantity (\ref{ratio}). For $n$ extra dimensions, the number of massive species with masses less than Hubble will be given by
\eq{Nn}{N\approx \left(\frac{H_*}{M}\right)^n \approx \left(\frac{H_*}{M_*}\right)^n\left(\frac{M_*}{M}\right)^n \approx \Upsilon^{n}\left(\frac{M_*}{M}\right)^n,} 
Furthermore, given that $\widetilde N = (M_{**}/M)^n = (M_{**}/H_*)^n(H_*/M)^n$ we arrive at the relation between the number of species that lower the effective cut-off during inflation $N$ with $\w N$ : 
\eq{ntilden}{N = \widetilde N \left(\frac{H_*}{M_{**}}\right)^n \equiv V_{**}\left(\frac{H_*}{M_{**}}\right)^n.}
As we shall see shortly, since $H_*/M_{**} < 1$ we have $N < \w N$, implying that in general we must also have $M_{**} < M_{*}$ as one can only cross additional mass thresholds from the scale of inflation to the scale of strong gravity. We note that (\ref{Nn}) immediately translates the uncertainty in the energy scale of inflation in terms of an intermediate compactification scale $M$ in units of $M_*$ through (\ref{vact}) and (\ref{ratio}):
\eq{vact2}{V_*^{1/4} \simeq  \, 3^{1/4}M_*\gamma^{1/2} = \,3^{1/4} \left(\frac{M}{\Upsilon M_*}\right)^{n/2}\Upsilon^{1/2}  \mpl,}
or equivalently
\begin{eqnarray}\label{vact3}
V_*^{1/4} &\simeq& \,
3^{1/4}\frac{M_{**}}{\mpl}\left(\frac{M_{**}}{H_*}\right)^{n/2}\Upsilon^{1/2}\, \mpl.
\end{eqnarray}
Now using (\ref{Hinf}), (\ref{Vinf}) and (\ref{ratio}), we have
\eq{VHrel}{V_*^{1/4}\simeq 3^{1/4}\, \Upsilon^{-1/2}\,H_*\,,}
so that equivalently
\eq{HMfun}{H_*\simeq M_{**}\, \,\Upsilon^{2/(n+2)}\,.}
It follows that $H_*$ is one to three orders of magnitude below the fundamental gravity scale $M_{**}$ for the range $0.001 \lesssim r_* \lesssim 0.1$. The ratio $H_*/M_*$ is of course fixed by (\ref{ratio}). Furthermore, we note that from (\ref{VHrel}) the energy scale of inflation is related to the scale $M_{**}$ by
\bea
\label{esf}
V_*^{1/4} &\simeq& 3^{1/4}\Upsilon^{2/(n+2) - 1/2}M_{**}
\eea 
which depending on the number of extra particles between $H_*$ and $M_{**}$ implies that $V_{*}^{1/4}$ can be greater than\footnote{For a large enough ratio $\w N/N$-- guaranteed for $n \geq 2$ through the hierarchy implied by (\ref{HMfun}).} $M_{**}$ (of the same order or an order of magnitude higher for $2\le n\le 6$), even though it is always less than the effective cut-off $M_*$ at the scale $H_*$ through (\ref{Vinf}). We note that this is never problematic even though $M_{**}$ is the cut-off induced by the underlying UV completion. This is because we remain in the perturbative regime with respect to corrections from the heavy states that UV complete the theory, which relies on derivatives being suppressed relative to this scale i.e. by the ratio $H_*/M_{**}$, guaranteed to be less than unity by (\ref{HMfun}). 

Furthermore, we stress that although extra dimensions (compactified at a scale below that of inflation) provide a natural context for the appearance of extra massive species, the relation (\ref{vact}) is also valid in a strictly four dimensional context and illustrates an irreducible uncertainty in our ability to infer a scale for inflation given our lack of knowledge of particle physics from collider energies up to the energy scale of inflation.   

\subsection{Large number of species in String Theory}

In the framework of string theory, the effective higher dimensional Planck mass $M_{**}$ is proportional to the fundamental string scale $M_s$, and eq.(\ref{usual}) becomes:
\eq{usualstring}{\mpl^2={1\over g_s^2}M_s^2V_{**},}
where $g_s$ is the string coupling and the internal volume $V_{**}$ is now given in string units. The corresponding number of species is then $\widetilde N=V_{**}/g_s^2$, which is fixed by the number of KK modes with mass lower than $M_s$ for $g_s\simeq{\cal O}(1)$, as is the case of D-branes where $g_s$ is given by the gauge coupling. Again, we distinguish $\widetilde N$, the number of KK modes below the effective cut-off around the compactification scale from $N$, the number of states with masses less than $H_*$. Note that the lower bound for the string scale of few TeV is consistent with a reheating temperature around above the electroweak scale (see discussion at the end of the previous section).

Apart from the possibility of having light KK modes of large extra dimensions, the fundamental gravity scale can be lowered due to a large number of species from hidden sectors (even coupled gravitationally to the Standard Model), or even from string excitations whose number increase exponentially with their mass. In the later case, the effective number of particle species which are not broad resonances, with width less than their mass is $\widetilde N\simeq 1/g_s^2$ \cite{ap,stringspecies}.

\section{(P)reheating}
The big bang begins shortly after inflation ends. The mechanism through which the inflaton dumps its energy density into the material content of the universe is known as reheating if this process occurs in thermal equilibrium, and preheating otherwise. During preheating, parametric resonance during the inflaton's final oscillations about its minimum results in bursts of particle production for any massive fields coupled to it (see \cite{reheat} and references therein for details on the points discussed here). The latter is a very out of equilibrium process and requires a subsequent period of thermalization. Since the primary mechanisms for generating parametric resonance have a purely particle physics origin, gravitational effects do not play any significant role and the mechanisms for preheating proceed as they do in the standard context regardless of the value of $M_*$. The exception being the special case of `geometric preheating', wherein the inflaton $\phi$ couples very weakly or has no direct couplings to a non-minimally coupled field $\chi$ with a non-minimal coupling parameter $\xi$. For this scenario, we first observe that on a background sourced by $\phi$ at the end of inflation, the mode functions for $\chi$ satisfy:
\eq{mfch}{\ddot\chi_k + 3H\dot\chi_k + \left(\frac{k^2}{a^2} + \xi R\right)\chi_k = 0.}
As the inflaton oscillates around its minimum, the scalar curvature $R = 12H^2 + 2\dot H$ oscillates as well. Upon time averaging we have the relation $\langle m_\phi^2\phi^2\rangle = \langle\dot\phi^2\rangle$ which implies $R \sim m_\phi^2\phi^2/M^2_*$, inducing an effective coupling to $\phi$ and which can produce parametric resonance for large enough $\xi$. By enhancing the strength of gravity, one enhances the effects of the geometric coupling term and thus widening the bands in which the Floquet index \cite{reheat} is positive, assisting parametric resonance non-linearly the more $M_*$ is reduced.

Reheating on the other hand is an equilibrium process that produces quanta of matter fields through one body decays such as $\phi \to \chi\chi$ or $\phi \to \bar\psi\psi$ where $\chi, \psi$ are scalar and fermionic quanta respectively. The interactions that can generate such decays are $\calL_{\phi\chi\chi} = \mu\phi\chi^2$ or $\calL_{\phi\bar\psi\psi} = y\phi\bar\psi\psi$, where $\mu$ has dimensions of mass and $y$ is dimensionless. In the limit $m^2_\phi \gg m^2_\psi , m^2_\chi$ the decay rates can be estimated as \cite{reheat}:
\bea
\Gamma_{\phi \to \chi\chi} &=&  \frac{\mu^2}{8\pi m_\phi} \\
\Gamma_{\phi \to \bar\psi\psi} &=& \frac{y^2 m_\phi}{8\pi}.
\eea 
Thermal equilibrium requires interactions to be efficient enough to equipartition all available states in phase space. In an expanding universe this necessitates $\Gamma_{tot} > H_*$. Hence, the maximum temperature reheating can occur at is implied by the condition $\Gamma_{tot} = H_* = \sqrt{\rho/(3M^2_*)}$. Assuming $g_*$ relativistic species in thermal equilibrium after reheating, we have
\eq{eqdist}{\rho = \frac{g_*\pi^2}{30}T^4,}
and therefore 
\eq{tmax}{T_{i} = \left(\frac{90}{g_*\pi^2}\right)^{1/4}\sqrt{\Gamma_{tot}M_*} \sim \left(\frac{90}{g_*\pi^2}\right)^{1/4}\frac{\sqrt{\Gamma_{tot}\mpl}}{N^{1/4}}.}
That is, the maximum temperature on can reheat is correspondingly reduced, consistent with our discussion in section III. We mention in closing that there are rich phenomenological possibilities in considering hidden sector fields produced in reheating as possible dark matter candidates in scenarios with many extra species, certain aspects of which have been studied in the multi-field inflationary context in \cite{Dvali:2009fw}\footnote{Also in the multi-field context are the references \cite{Nflation}. In these studies, no species \textit{below} the scale of inflation were considered, and so the strength of gravity at $H_*$ was taken to be the usual $\mpl$.}.

\section{Discussion}
It is commonly presumed that detection of a primordial tensor mode background would allow us to determine the (energy) scale of inflation in the context of single field inflation. The purpose of this note was to highlight the fact that instead, one can only \textit{infer} the (energy) scale of inflation from observations up to our ignorance of the scale $M_* = \mpl/\sqrt N$, the precise value for which depends on the spectrum of all universally coupled species with masses up to $H_*$, and for which field content of the standard model alone suggests an $N$ different from one though still of order unity. These observations raise the possibility that the energy scale for inflation can be significantly lowered by the presence of many gravitationally coupled species, an observation that has a particularly natural realization in extra dimensional scenarios, although is equally pertinent in a four-dimensional context. 

\section{Acknowledgements}
We wish to thank Robert Brandenberger, Cliff Burgess, Sergey Sibiryakov and Michael Trott for many informative discussions and comments on the draft. During an earlier portion of this collaboration, SP was supported by a Marie Curie Intra-European Fellowship of the European Community's 7'th Framework Program under contract number PIEF-GA-2011-302817.
\appendix

\section{Loop corrections to the graviton propagator}
\label{appendix}
As a concrete example to illustrate how gravity can become more strongly coupled at lower energies due to loops of matter fields, we reproduce the one loop correction to the graviton propagator on a flat Euclidean spacetime due to loops of a massive scalar field, as calculated by Capper in \cite{capper} (see also references therein). Beginning with the matter sector action\footnote{As we shall see shortly, the bare cosmological constant term is necessitated to ensure the satisfaction of the Slavnov-Taylor identities when expanding around flat space.}
\eq{mcs}{S = -\frac{2}{\kappa^2}\int d^4x\sqrt{g}\left(R - 2\Lambda\right) - \frac{1}{2}\int d^4x\sqrt{g}\left(g^{\mu\nu}\partial_\mu\p\partial_\nu\p  + m^2\p^2\right).}
It is particularly handy to define the tensor density
\eq{tddef}{\tilde g^{\mu\nu} = \sqrt g g^{\mu\nu},}
and define the graviton $\p^{\mu\nu}$ (note the departure from the usual  definition) as
\eq{gdef}{\tilde g^{\mu\nu} = \delta^{\mu\nu} + \kappa \p^{\mu\nu}} 
so that 
\eq{}{\sqrt g = det[g^{\mu\nu}]^{1/(d-2)} = exp\left[\frac{1}{d-2}Tr\,ln\left(\delta^{\mu\nu} + \kappa\p^{\mu\nu}  \right)\right].}
The matter Lagrangian can be expanded in powers of $\kappa$ as
\eq{intl}{\calL = \sum_{k=2} \kappa^{k-2}\calL_k}
with 
\eq{3pt}{\calL_3 = -\frac{1}{2}\p^{\mu\nu}\partial_\mu\p\partial_\nu\p + \frac{m^2}{2}\frac{\p^2\p^\mu_\mu}{d-2};~}
and
\eq{4pt}{\calL_4 = \frac{m^2\p^2}{d-2}\left[ -\p_{\mu\nu}\p^{\mu\nu} + \frac{\p^\mu_\mu\p^\nu_\nu}{d-2}\right]  }
which implies the cubic and quartic interaction vertices $V_{\alpha\beta}$ and $U_{\alpha\beta~\delta\gamma}$: 
\eq{verts}{V_{\alpha,\beta}(p,k_1,k_2) = -k_{1(\mu}k_{2\nu)} + \frac{m^2\delta_{\mu\nu}}{n-2};~~~~U_{\alpha\beta~\gamma\delta}(p_1,p_2,k_1,k_2) = \frac{m^2}{n-2}\left[-\delta_{\alpha(\gamma}\delta_{\beta(\delta)} + \frac{\delta_{\alpha\beta}\delta_{\gamma\delta}}{n-2}\right].}
\begin{figure}[H]
\label{vertices}
\epsfig{file=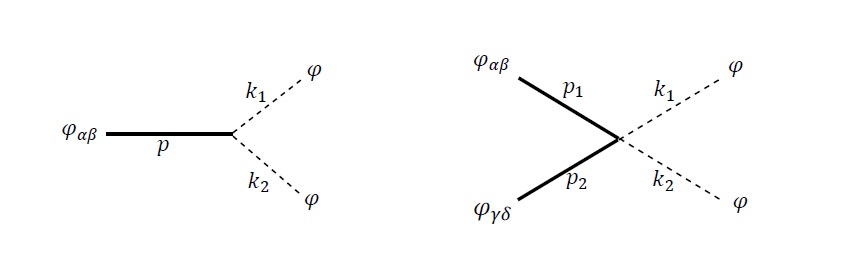,height=2in, width=6in}
\end{figure}
\noindent Supplemented with the massive scalar and graviton propagators (the latter obtained in de Donder gauge \cite{capper}):
\eq{props}{D(p) = \frac{-1}{p^2 + m^2};~~~~ D_{\alpha\beta~\mu\nu} = \frac{1}{2p^2}\left(\delta_{\alpha\mu}\delta_{\beta\nu} + \delta_{\alpha\nu}\delta_{\beta\mu} -\delta_{\alpha\beta}\delta_{\mu\nu}  \right)}
one can sum up all the self energy diagrams that contribute to the one loop corrected graviton propagator up to order $\kappa^2$: 
\begin{figure}[H]
\epsfig{file=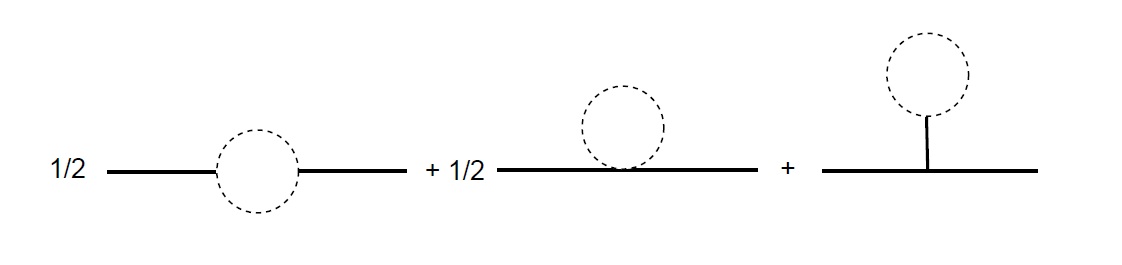,height=1.5in,width=6.5in}
\caption{\label{gravprop} Diagrams contributing to the one loop correction to the graviton propagator.}
\end{figure}
\noindent Formally, the loop corrected propagator $Q_{\nu\sigma\mu\lambda}$ is obtained by inserting the sum of all self energy diagrams: 
\eq{propcorr}{Q_{\nu\sigma\mu\lambda} = \kappa^2 D_{\nu\sigma}^{~~~\alpha\beta}T_{\alpha\beta\theta\tau}D^{\theta\tau}_{~~~~\mu\lambda},}
which evaluates to
\begin{eqnarray}
\label{ans}
Q_{\nu\sigma\mu\lambda} &=& \frac{\kappa^2}{p^4}\Biggl\{T_1\left[p_\nu p_\sigma p_\mu p_\lambda - p^2\delta_{\mu\lambda}p_\nu p_\sigma - p^2\delta_{\nu\sigma}p_\mu p_\lambda  + p^4\delta_{\nu\sigma}\delta_{\mu\lambda}\right] \\ \nn &+& T_3\left[p^4\delta_{\mu\nu}\delta_{\sigma\lambda} + p^4\delta_{\nu\lambda}\delta_{\sigma\mu} - 2p^4\delta_{\nu\sigma}\delta_{\mu\lambda} + 2p^2\delta_{\mu\lambda}p_\nu p_\sigma + 2p^2\delta_{\nu\sigma}p_\mu p_\lambda - 4p^2 p_{(\nu}\delta_{\sigma)(\mu}p_{\lambda)}\right]\Biggr\}
\end{eqnarray}
where $2\omega = d$ is the parameter through which we implement dimensional regularization, $\mu$ is an arbitrary renormalization scale and where 
\begin{eqnarray}
T_1 &=& m^{2(\omega-2)}\frac{\mu^{2(2-\omega)}}{2(4\pi)^\omega}\Gamma(2-\omega)\Biggl[\frac{1}{3}~_3F_2\left(2-\omega,1,3;2,5/2;-\frac{p^2}{4m^2}\right)\\ \nn &-& \frac{1}{2}~_3F_2\left(2-\omega,4,1;5/2,3,-\frac{p^2}{4m^2}\right) + \frac{1}{5}~_3F_2\left(2-\omega,5,1;3,7/2,-\frac{p^2}{4m^2}\right)\Biggr]
\end{eqnarray}
\begin{eqnarray}
\label{t2}
T_3 = m^{2\omega}\frac{\mu^{2(2-\omega)}}{8(4\pi)^\omega}\frac{\Gamma(2-\omega)}{p^4(\omega-1)}\Biggl[\frac{1}{1-\omega}+\frac{1}{\omega}~_2F_1\left(-\omega,1;3/2;-\frac{p^2}{4m^2}\right)\Biggr]
\end{eqnarray}
Some remarks are in order at this stage. There are five possible terms with the right tensor structure that could contribute to $T_{\alpha\beta\theta\tau}$, three of whose coefficients can be eliminated by repeated applications of the Ward identities. Doing so, results in (\ref{ans})\footnote{Furthermore, the addition of the bare cosmological term in (\ref{mcs}) was necessitated by the non-vanishing tadpole in Fig. \ref{gravprop}, which we have to cancel in order to consistently expand around flat space-- the remaining diagrams thus resulting in (\ref{ans})-(\ref{t2}).}, for which it is easily verified that $p^\nu Q_{\nu\sigma\mu\lambda} = 0$. Having dimensionally regularized the loop integrals, we find the usual $1/(2-\omega)$ poles which require appropriate counterterms to subtract the divergences. These are given (in Lorentzian signature) as
\eq{counter}{\calL_{c.t.} = -\frac{\sqrt{-g}}{16\pi^2}\frac{1}{(2-\omega)}\left\{ \frac{m^4}{4} + \frac{m^2}{12}R + \frac{1}{120}\left[\frac{R^2}{2} + R^{\mu\nu}R_{\mu\nu}\right] \right\} } 
Given the asymptotic behaviour of the Hypergeometric and Gamma functions, it is straightforward to evaluate the remaining (finite) part of $Q_{\nu\sigma\mu\lambda}(p)$ in the limit $p^2 \gg m^2$, where we find (suppressing tensor structure), the expected conformal limit for $N$ minimally coupled massive scalars (cf. the result of (\ref{climit}) plugged in to (\ref{propcorr}) and recalling that the usual definition of the metric perturbation $h_{\mu\nu} := \kappa\p_{\mu\nu}$)
\eq{climit2}{\boxed{\lim_{p^2/m^2 \to \infty} \|Q_{\nu\sigma\mu\lambda}\| \sim \kappa^2 \frac{N}{16\pi^2} {\rm log} \left(p^2/\mu^2\right)}}

\section{Note on universally coupled massive species}
\label{note}
\begin{figure}[h]
\begin{center}
\epsfig{file=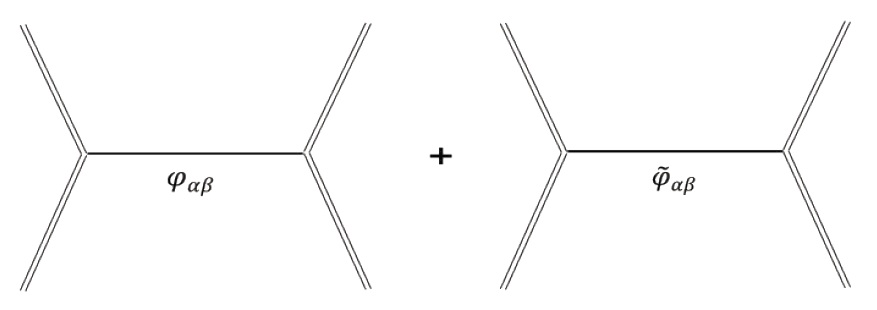,height=1.6in, width=4.2in}
\end{center}
\caption{\label{massiveKK} Correction to the gravitational interaction between conserved sources from crossing a massive KK graviton $\widetilde\p_{\alpha\beta}$.}
\end{figure}

We wish to demonstrate that although all massive species contribute to lowering the effective scale of strong gravity, that the threshold effects imparted by universally coupled species (e.g. massive KK gravitons) differs qualitatively from that of massive, non-universally coupled species. Whereas the former changes the strength of gravity by an order one effect as you cross the threshold $M$, the latter only changes the strength of gravity by a factor of $M^2/\mpl^2$, even as both contribute equally to lowering the scale at which strong gravity effects become important. This can be seen by considering the correction to the graviton propagator from loops of $N$ ordinary massive fields in the limit of high momentum transfer $p^2 \gg M^2$:
\eq{gpc}{\frac{1}{\mpl^2 p^2} + \frac{1}{\mpl^4 p^4}\,\frac{N p^4}{16\pi^2} {\rm log} \left(p^2/\mu^2\right) + ...}
Repeated insertions of self energy diagrams would result in a geometric series in the above of which we can compute the (leading log) resummed propagator:
\eq{resum}{\frac{1}{\mpl^2 p^2\left[1 -  \frac{N p^2}{\mpl^2}{\rm log} \left(p^2/\mu^2\right)\right]}}
Clearly from (\ref{gpc}) we can infer the usual scale of strong gravity $M_{**} = \mpl/\sqrt N$, however, if we wanted to interpret (\ref{resum}) as a strengthening of gravity as you cross the mass threshold $M$, we see that the effect is not significant until you come very close to $M_{**}$. That is, in the regime $M^2 \ll p^2 \ll \mpl^2/N$, the usual Newton's force would result with $G_N = \mpl^{-2}$. This can also be appreciated by directly computing the one loop corrected Newton's potential between two conserved sources. Using (\ref{ans}) to (\ref{t2}), we can compute the gravitational field generated by the source $T_{\lambda\beta}$ given by (in the notation of appendix A):
\eq{csource}{\p_{\mu\nu}(x) = \int \frac{d^Dp}{(2\pi)^{D/2}}\, e^{ip\cdot x}\left[D_{\mu\nu}^{~~~\lambda\beta}(p) +  Q_{\mu\nu}^{~~~\lambda\beta}(p)\right]T_{\lambda\beta}(p).}
Consider the energy momentum tensor of a point mass at rest, given by
\eq{emtpm}{T_{\lambda\beta}(x) = m\delta^{D-1}(\vec x)\delta^0_\lambda\delta^0_\beta~~;~~ T_{\lambda\beta}(p) = \frac{2\pi m}{(2\pi)^{D/2}}\delta(p^0)\delta^0_\lambda\delta^0_\beta }
so that
\eq{csource2}{\p_{\mu\nu}(x) = 2\pi m\int \frac{d^{D-1}p}{(2\pi)^{D}}\, e^{i\vec p\cdot \vec x}\left[D_{\mu\nu}^{~~~00}(p) + Q_{\mu\nu}^{~~~00}(p)\right]|_{p^0 = 0}.}
One can revert to the traditional dimensionless metric perturbation $h^{\mu\nu}$ via the relation (\ref{tddef}):
\eq{revert}{h^{\mu\nu} = -\frac{1}{\mpl}\left[\p^{\mu\nu} - \frac{\p^\lambda_\lambda}{D-2}\delta^{\mu\nu}\right].} 
Ignoring the second term and using the expression for the propagator in de Donder gauge
\eq{}{\nn D_{\alpha\beta~\mu\nu} = \frac{1}{2p^2}\left(\delta_{\alpha\mu}\delta_{\beta\nu} + \delta_{\alpha\nu}\delta_{\beta\mu} -\delta_{\alpha\beta}\delta_{\mu\nu}\right)}
we find 
\eq{newton}{\p_{ij} = -\frac{m}{8\pi r}\delta_{ij},~~\p_{00} = \frac{m}{8\pi r},}
which through (\ref{revert}) implies the usual Newton's potential. From (\ref{ans}) to (\ref{t2}) the one loop correction results in three possible finite contributions to the integrand (suppressing tensor structure and factors of order unity)-- 
\eq{}{16\pi^2 \|Q\| \sim \frac{1}{\mpl^2} {\rm log} \left(M^2/\mu^2\right) + \frac{1}{\vec p^2}\frac{M^2}{\mpl^2} {\rm log} \left(M^2/\mu^2\right) + \frac{M^2}{\vec p^4}\frac{M^2}{\mpl^2} {\rm log} \left(M^2/\mu^2\right) }
The first two terms contribute a contact (delta function) term and a 
correction to the usual Newton's potential respectively. The third term contributes linear and logarithmic corrections that depend on $r M$, which are only to be understood in the regime where they are small corrections that get completed by higher order terms in the perturbative expansion such that the total amplitude satisfies the usual decoupling requirements (i.e. that the corrections from insertion of self energy diagrams of massive particles vanish in the long wavelength limit). Clearly the latter two terms have contributions that are $M^2/\mpl^2$ suppressed as per the usual expectation \cite{Donoghue, Burgess}. Therefore crossing any particular threshold scale $M$ does not result in a significant strengthening of gravity around that scale, whose summed effects accumulate only close to the scale $M_{**}$ as per our discussion around (\ref{resum}).
\begin{figure}[H]
\begin{center}
\epsfig{file=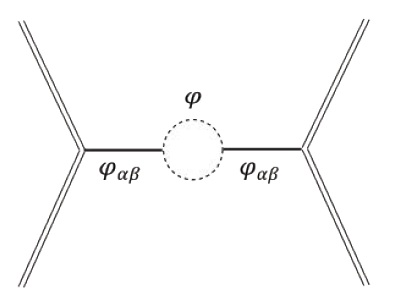,height=1.6in, width=2in}
\end{center}
\caption{\label{massive} Correction to the gravitational interaction between conserved sources (denoted by a double line) from crossing the mass threshold of a massive particle.}
\end{figure} 
\noindent The situation for massive KK gravitons is markedly different however as these resonances couple universally to all conserved sources and can therefore correct their gravitational interactions at tree level through the diagrams in fig. \ref{massiveKK}.   
\noindent Lets say there were $n$ KK resonances with mass $M$, then it is straightforward to calculate the correction to the gravitational interaction--
\eq{kkgrav}{\frac{1}{\mpl^2 p^2} \to \frac{1}{\mpl^2 p^2} + \frac{n}{\mpl^2 (p^2+M^2)}.}
Unlike the case for (\ref{resum}), we see that the regime $M^2 \ll p^2 \ll \mpl^2/n$ the strength of gravity is modified immediately above $p = M$ as 
\eq{kkgravthresh}{\frac{1}{\mpl^2 p^2} + \frac{n}{\mpl^2p^2 \left(1+M^2/p^2\right)} \to \frac{n+1}{\mpl^2p^2},}
that is, as each threshold corresponding to a massive KK resonance is crossed, the effective strength of gravity increases immediately as $\mpl^2/n \to \mpl^2/(n+1)$, where $n$ counts the number of massive species that contribute to the tree level diagram fig. \ref{massiveKK}. We also note that effective interactions of gravitational strength can also be generated from universally coupled species such as the Higgs through higher dimensional operators of the form
\eq{dim6}{\Delta\calL_{\rm eff} \sim c_1\frac{H^\dag H}{\mpl^2}\partial_\mu\p\partial^\mu\p + c_2\frac{H^\dag H}{\mpl^2}\bar\psi\slashed\partial\psi\sim  c_{\{1,2\}}\frac{H^\dag H}{\mpl^2}T^\mu_\mu,}
where the $c_i$ are generic Wilson coefficients. Expanding the singlet operator $H^\dag H$ around some vev $v$ as $H^\dag H = v^2 + 2 v h$  generates a vertex that contributes another channel to the diagram fig.\ref{massiveKK} 
\eq{d5}{\Delta\calL_{\rm eff}\sim c_i\frac{v\,h}{\mpl^2}T^\mu_\mu}
so that above the effective Higgs mass $m_H$, in addition to the usual massless graviton exchange, one mediates an extra gravitational strength force 
\eq{higgscont}{\frac{1}{\mpl^2 p^2} \to \frac{1}{\mpl^2 p^2} + \frac{g_i^2}{\mpl^2 (p^2+m_H^2)},}
with 
\eq{gval}{g_i^2 := c^2_i v^2/\mpl^2.}
Just as in the case for massive KK gravitons, one finds that in the regime $m_H^2 \ll p^2 \ll \mpl^2$ the effective strength of gravity is enhanced as
\eq{Higgsuniv}{\mpl^2 \to \frac{\mpl^2}{\left(1 + \sum_ig_i^2\right)}.}
We realize that perhaps this shouldn't be too surprising, as one can always field redefine the operators (\ref{dim6}) via the trace of the (two derivative) background equations of motion \cite{weinberg} $R = -T^\mu_\mu/\mpl^2$ so that the effective interactions are equivalent to
\eq{nmint}{\Delta\calL_{\rm eff} \sim c_{\{1,2\}}H^\dag H R.}
Therefore we see that it is not just the Higgs, but any non-minimally coupled massive scalar\footnote{Order unity non-minimal couplings will generically be generated through renormalization group (RG) flow for the singlet component of the Higgs and any other massive scalars present in the early universe \cite{BD}.}
that can enhance the effective strength of gravity as (\ref{Higgsuniv}), where the Wilson coefficients $c_i$ are now replaced with non-minimal couplings $\xi_i$. We stress that such additional effective interactions generate extra gravitational strength scalar forces that violate the equivalence principle at very high energies, but that this is not in conflict with any presently accessible observations. To briefly recap, the physical basis of the differing effects of non-universally coupled compared to universally coupled massive species on the strength of gravity as you cross each threshold can be readily understood through the former being a loop effect and the latter being a tree level effect cf. figs (\ref{massive}, \ref{massiveKK}).

One can complement this understanding of how regular massive species do not affect the strength of gravity as seen by curvature quanta during inflation through the simple exercise of computing the action for $\calR$ given the fact that the effects of massive particles in fig. \ref{massive} can be reproduced by the effective action
\eq{hdact}{S = \frac{M_R^2}{2}\int d^4x\sqrt{-g}\,R + \int d^4x\sqrt{-g}\left[c_1 R^2 + c_2 R^{\mu\nu}R_{\mu\nu} \right] + \int d^4x\sqrt{-g}\calL\,[\phi_0]}
where the last term in the above is the action for the background that sources the quasi de Sitter phase, and the subscript on $\phi_0$ anticipates that we will work in a gauge where all inflaton fluctuations have been gauged away. The dimensionfull prefactor $M^2_R$ emphasizes that the above is the net result of having subtracted off the usual divergences that result after integrating out the fields in question and are fixed by renormalization conditions at some particular scale, necessitating the introduction of a bare cosmological constant term that is cancelled by tadpole contributions. One can straightforwardly deduce that the effect of the higher curvature terms will be to modify the usual action for the comoving curavture perturbation from
\eq{usual}{S_2 = \mpl^2\int d^4x \,a^3\epsilon\left[\dot\calR^2 - \frac{(\partial\calR)^2}{a^2}\right] }
to
\eq{nonusual}{S_2 = M_*^2\int d^4x \,a^3\epsilon\left[\frac{\dot\calR^2}{c_s^2} - \frac{(\partial\calR)^2}{a^2} + \frac{\lambda}{M_*^2}\frac{(\partial\calR)^4}{a^4} + ... \right] }
where we see from (\ref{hdact}) that the corrections from the curvature squared terms to the quadratic action for $\calR$ must have two derivatives acting on background quantities\footnote{In full generality we allow for the generation of a non-trivial the adiabatic sound speed $c_s < 1$ as well as higher spatial derivative terms from the curvature squared corrections.}, so that for example, the operator $(\partial\calR)^2$ gets a correction of the form
\eq{opcorr}{\Delta \calL_2 \sim c_{1,2}M_R^2\frac{(\partial\calR)^2}{a^2} \frac{H^2}{M_R^2}}
where $c_1, c_2 \sim \mathcal O(N)$ so that the overall dimensionful coefficient of the quadratic action (\ref{nonusual}) becomes
\eq{mstarhd}{M_*^2 = M_R^2\left(1 - \frac{\tilde c N H^2}{16\pi^2M_R^2}  \right) }
where $\tilde c$ is some positive number (by unitarity) of order unity. Furthermore, we note that integrating out massive particles can only additively renormalize Newton's constant such that at the scale of inflation, $M_R^2(\mu) = M^2(\mu_{\rm ref}) + \sum_i d_i m^2_i\,{\rm log}(\mu^2/\mu_{\rm ref}^2)$ where the $d_i$ are spin dependent weights each suppressed by a factor of $1/(16\pi^2)$. Imposing the renormalization conditions at macroscopic scales, $M^2(\mu_{\rm ref}) = \mpl^2$ so that 
\eq{Mr}{M^2_R(\mu) = \mpl^2\left(1 +  \sum_i d_i \frac{m^2_i}{\mpl^2}\,{\rm log}(\mu^2/\mu_{\rm ref}^2)\right) \approx \mpl^2}
within the domain of validity of our approximation. Therefore in the context of (\ref{mstarhd}), we see that unless $H^2 \sim \mpl^2/\sqrt N$, the correction to the effective strength of gravity as seen by the curvature perturbations is negligible and $M_* \sim \mpl$ for curvature quanta above the scale of any massive (non-universally coupled) species.  
\section{Moduli dynamics during inflation}
\label{moduli}
In this appendix, we consider the dynamics of moduli fields during inflation in the case where $H_* > \mu_c$, where $\mu_c$ is the characteristic mass of the moduli. We wish to show that are regimes where the moduli are effectively frozen even if they are dynamically displaced off their minima, allowing us to treat the extra dimensional volume as effectively constant during inflation. We begin by considering the following D-dimensional action
\eq{ddact}{S = \frac{M_*^{D-2}}{2}\int d^Dx\sqrt{-G}\,R^{(D)} - \int d^Dx\sqrt{-G}\left[\frac{1}{2}(\partial\phi)^2 + V(\phi)\right]}
where $R^{(D)}$ is the D-dimensional Ricci scalar constructed out of the metric $G_{AB}$. We now consider a $D = 4 + n$ decomposition with Greek indices denoting four non-compact coordinates and lower case Latin indices denoting $n$ periodic extra dimensional coordinates that range from $0 \leq y^a \leq 2\pi R$ and where we furthermore presume the metric tensor $G_{AB}$ to be defined through the block diagonal form
\eq{bdm}{ds^2 = g_{\mu\nu}(x)dx^\mu dx^\nu + e^{2\omega(x)}\gamma_{ab}dy^a dy^b.}
If we assume $\gamma_{ab} = \delta_{ab}$, and if no other quantity depends on the $y^a$ then 
\eq{metG}{\sqrt{-G} = \sqrt{-g}\,e^{n\omega}}
and
\eq{Riccidec}{R^{(D)}[G] = R^{(4)}[g] - 2n e^{-\omega}g^{\mu\nu}\nabla_\mu\nabla_\nu e^\omega - n(n-1)g^{\mu\nu}\nabla_\mu \omega\nabla_\nu \omega.}
Substituting into the above, integrating over the compact dimensions and after a few integrations by part results in
\eq{ddact2}{S = \frac{M_*^{D-2}}{2}V_n\int d^4x\sqrt{-g}\,e^{n\omega}\left[R^{(D)} + n(n-1)(\partial\omega)^2\right] - V_n\int d^4x\sqrt{-g}\,e^{n\omega}\left[\frac{1}{2}(\partial\phi)^2 + V(\phi)\right],}
where $V_n := \int d^ny$. To bring the action above into the usual Einstein Hilbert form, we further make the conformal transformation
\eq{conft}{g_{\mu\nu} = e^{-n\omega}\w g_{\mu\nu}}
so that finally, the action becomes
\eq{ddactf}{S = \frac{\mpl^2}{2}\int d^4x\sqrt{-\w g}\,\left[R^{(D)}[\w g] - \frac{n}{2}(n+2)(\w\partial\omega)^2 - U(\omega)\right] - \int d^4x\sqrt{-\w g}\,\left[\frac{1}{2}(\w\partial\w\phi)^2 + e^{-2\omega} V(\w\phi)\right]}
where $\mpl^2 := M_{*}^{(D-2)}V_n$ and $\w\phi := \sqrt V_n \phi$, and where we have explicitly introduced by hand the Einstein frame  potential $U$ that is responsible for stabilizing the $\omega$ modulus\footnote{For a review of mechanisms to stabilize K\"ahler moduli in the context of type II and heterotic string theory, see \cite{moduli}. See also \cite{free} for complementary approaches utilizing the energetics of the string free energy around enhanced symmetry points.}. Restricting ourselves to spatially homogeneous solutions, we make the metric ansatz
\eq{4dma}{ds^2 = -dt^2 + e^{2\lambda}\delta_{ij}dx^idx^j}
and find the Einstein constraint equation (dropping the tilde's on $\phi$ in what follows),
\eq{00}{3\dot\lambda^2 = \frac{n(n+2)}{4}\dot\omega^2 + \frac{U(\omega)}{2} + \frac{\dot{\phi^2}  }{2\mpl^2} + \frac{e^{-2\omega}}{\mpl^2}V}   
the equation of motion 
\eq{eomphi}{\ddot{\phi} + 3\dot\lambda\dot{\phi} + e^{-2\omega}V_{,\phi} = 0}
and the Friedmann equations
\eq{F1}{\ddot\lambda + \frac{n(n+2)}{4}\dot\omega^2 + \frac{\dot{\phi^2}  }{2\mpl^2} = 0}
\eq{F2}{\ddot\omega + 3\dot\lambda\dot\omega + \frac{U_{,\omega}}{n(n+2)} - \frac{4e^{-2\omega}}{n(n+2)\mpl^2}V = 0.} 
We now ask, is there a solution such that the background inflates, i.e. $-\dot H/H^2 \ll 1$? From (\ref{F1}), this clearly requires
\eq{sr1}{-\frac{\dot H}{H^2} = - \ddot\lambda/\dot\lambda^2 = \frac{3}{3\dot\lambda^2}\left[  \frac{n(n+2)}{4}\dot\omega^2 + \frac{\dot{\phi^2}  }{2\mpl^2}\right] \ll 1}
where the denominator on the rhs is given by (\ref{00}). Defining the total kinetic and potential energies
\eq{KTPT}{K_T :=  \mpl^2\frac{n(n+2)}{4}\dot\omega^2 + \frac{\dot{\phi^2}  }{2};~~~~V_T := \mpl^2\frac{U(\omega)}{2} + e^{-2\omega}V(\phi),}
we find that in general
\eq{hdex}{\epsilon = -\frac{\dot H}{H^2} = \frac{3K_T}{K_T + V_T}}
so that $\epsilon \ll 1$ consistently if $K_T/V_T \ll 1$, i.e. we require the potential terms dominate the kinetic terms in the energy density, as per usual. This can be re-expressed using (\ref{eomphi}) and (\ref{F2}) as a condition on the potential if one neglects the $\ddot\phi$ and $\ddot\omega$ terms, which is only possible if
\eq{finanssr}{ \frac{1}{n(n+2)}\frac{V^2_{T,\omega}}{V_T^2} + \frac{\mpl^2}{2}\frac{V^2_{T,\phi}}{V_T^2} \ll 1,}
where the second term above is the usual $\epsilon$ parameter. We note from the definition (\ref{KTPT}) that requiring $U(\omega)e^{2\omega}/V(\phi) \ll 1$ (which is consistent with requiring inflation being above the effective compactification scale $H_* \gg \mu_c$ over a broad regime) results in 
\eq{srcfin}{\frac{4}{n(n+2)} + \frac{\mpl^2}{2}\frac{V'(\phi)^2}{V(\phi)^2} \ll 1,}
so that in addition we also need a sufficient number of extra dimensions in order that the second numerical factor can also be neglected\footnote{With six extra dimensions for example, this factor equals $1/12$. One may be tempted to redefine $\omega$ in (\ref{ddactf}) to absorb the factor of $n(n+2)$, however this appropriately rescales the arguments of the potential energy contributions such that (\ref{srcfin}) also results for the redefined variables.}. One can understand why this is so-- we see from (\ref{ddactf}) that the for the canonically normalized variable $\w\omega = \sqrt{n(n+2)}\omega$, the inflationary potential becomes $e^{-2\w\omega/\sqrt{n(n+2)}}V(\phi)$, therefore one flattens the $\w\omega$ dependence of this contribution to the total potential more and more the greater $n$ is, eventually allowing for the `slow-roll' condition on the moduli field to be satisfied.

Therefore although it is not generally true that we can have inflation without the moduli undergoing non-trivial excursions we see that in certain limits, this can indeed be accomplished consistently allowing us to treat them as effectively fixed even though they are displaced from their minima. In this situation, the characteristic masses of the associated KK states remains almost constant during inflation and the usual analysis can be implemented.

\end{document}